\documentclass[referee,pdflatex,sn-mathphys-num]{sn-jnl}

\usepackage{graphicx}%
\usepackage{multirow}%
\usepackage{amsmath,amssymb,amsfonts,amsthm}%
\usepackage{mathrsfs}%
\usepackage[title]{appendix}%
\usepackage{xcolor}%
\usepackage{textcomp}%
\usepackage{manyfoot}%
\usepackage{booktabs}%
\usepackage{algorithm}%
\usepackage{algorithmicx}%
\usepackage{algpseudocode}%
\usepackage{listings}%

\usepackage{physics}
\usepackage{siunitx}

\usepackage{braket}
\usepackage{gensymb}
\usepackage{comment}
\usepackage{xcolor}
\usepackage[normalem]{ulem}

\usepackage{booktabs, mathtools}
\usepackage{helvet}

\usepackage[T1]{fontenc}
\usepackage{anyfontsize}
\usepackage[left]{lineno}

\theoremstyle{thmstyleone}%
\theoremstyle{thmstyletwo}%

\theoremstyle{thmstylethree}%

\begin{document}

\newcommand{\ben}[1]{{\color{olive}#1}}


\title[Article Title]{Rapid high-temperature initialisation and readout of spins in silicon with 10 THz photons}

\author[1, 2, 3]{\fnm{Aidan G.} \sur{McConnell}}

\author[4]{\fnm{Nils} \sur{Dessmann}}

\author[3, 5]{\fnm{Wojciech} \sur{Adamczyk}}

\author[6]{\fnm{Benedict N.} \sur{Murdin}}

\author[2, 3, 7]{\fnm{Lorenzo} \sur{Amato}}

\author[8]{\fnm{Nikolay V.} \sur{Abrosimov}}

\author[9]{\fnm{Sergey G.} \sur{Pavlov}}

\author*[1, 2, 3, 10]{\fnm{Gabriel} \sur{Aeppli}}\email{aepplig@ethz.ch}

\author*[1]{\fnm{Guy} \sur{Matmon}}\email{guy.matmon@psi.ch}

\affil[1]{\orgdiv{PSI Center for Photon Science}, \orgname{Paul Scherrer Institute}, \orgaddress{\city{Villigen PSI}, \postcode{CH-5232}, \country{Switzerland}}}

\affil[2]{\orgdiv{Laboratory for Solid State Physics}, \orgname{ETH Zurich}, \orgaddress{\city{Zurich}, \postcode{CH-8093}, \country{Switzerland}}}

\affil[3]{\orgdiv{Quantum Center}, \orgname{ETH Zurich}, \orgaddress{\city{Zurich}, \postcode{CH-8093}, \country{Switzerland}}}

\affil[4]{\orgdiv{Institute for Molecules and Materials, FELIX Laboratory}, \orgname{Radboud University}, \orgaddress{\city{Nijmegen}, \country{The Netherlands}}}

\affil[5]{\orgdiv{Institute for Quantum Electronics}, \orgname{ETH Zurich}, \orgaddress{\city{Zurich}, \postcode{CH-8093}, \country{Switzerland}}}

\affil[6]{\orgdiv{Advanced Technology Institute and Department of Physics}, \orgname{University of Surrey}, \orgaddress{\city{Guildford}, \postcode{GU2 7XH}, \country{UK}}}

\affil[7]{\orgdiv{PSI Center for Scientific Computing, Theory and Data}, \orgname{Paul Scherrer Institute}, \orgaddress{\city{Villigen PSI}, \postcode{CH-5232}, \country{Switzerland}}}

\affil[8]{\orgname{Leibniz-Institut für Kristallzüchtung (IKZ)}, \orgaddress{\city{12489 Berlin}, \country{Germany}}}

\affil[9]{\orgdiv{Institute of Space Research, German Aerospace Center (DLR)}, \orgaddress{\city{12489 Berlin}, \country{Germany}}}

\affil[10]{\orgdiv{Institute of Physics}, \orgname{EPF Lausanne}, \orgaddress{\city{Lausanne}, \postcode{CH-1015}, \country{Switzerland}}}

\abstract{
Each cycle of a quantum computation requires a quantum state initialisation. For semiconductor-based quantum platforms, initialisation is typically performed via slow microwave processes and usually requires cooling to temperatures where only the lowest quantum level is occupied. In silicon, boron atoms are the most common impurities. They bind holes in orbitals including an effective spin-3/2 ground state as well as excited states analogous to the Rydberg series for hydrogen. Here we show that initialisation temperature demands may be relaxed and speeds increased over a thousand-fold by importing, from atomic physics, the procedure of optical pumping via excited orbital states to preferentially occupy a target ground state spin. Spin relaxation within the orbital ground state of unstrained silicon is too fast to measure for conventional pulsed microwave technology, except at temperatures below \SI{2}{\kelvin}, implying a need not only for fast state preparation but also fast state readout. Circularly polarised $\sim$\SI{10}{\tera\hertz} photon pulses from a free electron laser meet both needs at temperatures above \SI{3}{\kelvin}: a \SI{9}{\pico\second} pulse enhances the population of one spin eigenstate for the ``1s''-like ground state orbital, and the second interrogates this imbalance in spin population. Using parameters given by our data, we calculate that it should be possible to initialise 99\% of spins for boron in strained silicon within \SI{250}{\pico\second} at \SI{3}{\kelvin}. The speedup of both state preparation and measurement gained for THz rather than microwave photons should be explored for the many other solid state quantum systems hosting THz excitations potentially useful as intermediate states. 
}

\maketitle

Optical pumping transformed atomic physics by enabling high fidelity quantum state preparation \cite{Kastler1950, Kastler_NobelLecture_1966, Happer1972}. It underpins technologies ranging from atomic clocks and trapped-ion quantum computing to brain and cardiac imaging \cite{Livanov1981_OPM_MCG, Boto2018}. Here we demonstrate the same type of state preparation for atoms not in vacuum, but for the common acceptor, boron, in silicon. 

Spin qubits in semiconductors remain among the most promising candidates for quantum information processing due to their long coherence times \cite{Tyryshkin2012, Stano2022} and compatibility with the fabrication infrastructure of conventional information technology \cite{Burkard2023, Zwanenburg2013, Constantinou2024}. Acceptor-bound holes in silicon provide particularly attractive spin qubits: they feature strong spin-orbit coupling, allowing all-electrical control and enabling simpler, more scalable architectures than electron-based schemes \cite{Salfi_2016_2, Salfi2016, Zhang2023, Heijden2018, Kobayashi2021, Hendrickx2020}. 

A major challenge for both electron and hole-based qubit platforms is state initialisation, which must be repeated often to build up the results of an actual quantum computation. Long spin lifetimes are desirable for quantum memory and computation, but make conventional thermal initialisation slow: at millikelvin temperatures and modest magnetic field splitting, seconds to minutes are required for spins to equilibrate. The limitation can be partially alleviated through enhancement of the coupling between spin states and the environment, temporarily shortening the qubit lifetime ($T_{1}$) and thereby mediating rapid thermalisation. In practice, this is achieved by controlling the electric field near qubits using gate electrodes. However, as the method does not increase the spin splitting $\Delta$, high-fidelity initialisation still requires millikelvin temperatures $k_{B}T\ll\Delta$. Recently there have been numerous more elaborate examples of initialisation, manipulation and readout which function at higher cryostat temperatures \cite{Yang2020, Petit2020, Camenzind2022, Huang2024, Geng2025}. However, initialisation remains relatively slow, lasting at least microseconds in single dopant and gated silicon devices \cite{Stano2022, Burkard2023, Watson2015, Johnson2022, Reiner2024}.

\begin{figure}[b]
\centering
\includegraphics[width=1.0\linewidth]{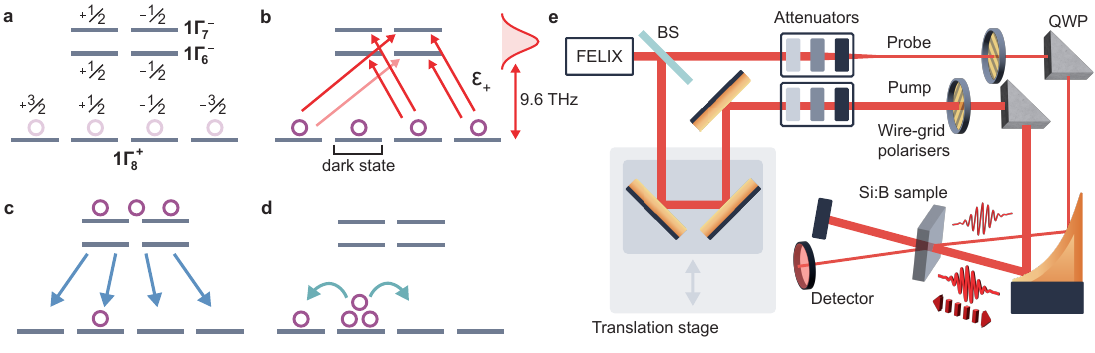}
\caption{
\textbf{Excitation--relaxation cycle and experimental schematic.} 
\textbf{a-d} Each step in an initialisation cycle, described in the main text. \textbf{b} shows the allowed transitions from the $1\Gamma_8^+$ ground state quartet to the $1\Gamma_{6,7}^{-}$ excited doublets for $\epsilon_{+}$ light propagating along $\langle111\rangle$. The faint red line indicates a near-zero transition rate \cite{Bhattacharjee1972}. The laser spectrum overlaps differently with the excited states. \textbf{e} A beam-splitter (BS) splits the linearly-polarised FELIX pulse into pump and probe beams. The pump passes a translation stage to control the relative arrival time between the two pulses. Both beams then pass wire-grid polarisers that can be independently rotated to control the final polarisations incident on the silicon prism quarter wave plates (QWP). Both beams coincide on the \SI{2.9}{\kelvin} boron-doped silicon. The angle of incidence differs by $\sim$\SI{5}{\degree} to filter the pump light out of the detector channel. 
}
\label{fig:cycle_schematic}
\end{figure}

Here we propose a more straightforward and faster scheme for high temperature qubit operation in silicon, namely optical pumping and readout of hole spins bound to boron acceptors in silicon, using circularly polarised $\sim$\SI{10}{\tera\hertz} pulses. 
By exciting the “1s”-like ground state \cite{Kittel1954, Kohn1957, Baldereschi1973} to higher-lying hydrogenic orbitals using a $\sim$\SI{9}{\pico\second} laser pulse and relying on fast, phonon-mediated relaxation, we increase the population of the target spin by $\sim$12\%. The transition energy is orders of magnitude larger than the spin splitting $\Delta$, trivially enabling initialisation at high temperatures. 

We then put this method to immediate practical use to measure the spin-lattice relaxation time for the bound holes in a regime not accessible to ordinary pulsed microwave spectroscopy, without ever using a magnetic field. Finally, we describe calculations showing that our THz method should reach $>$99\% initialisation within hundreds of picoseconds, at least three orders of magnitude faster than in similar dopant-bound qubits \cite{Watson2015, Johnson2022, Yang2009, Lo2015, Reiner2024, Burkard2023, Gritsch2025}. 

\section*{Optical pumping}

\subsection*{Experiment}

Just as in Kastler's original proposal \cite{Kastler1950}, we employ a circularly polarised pump beam tuned to a specific optical transition with selection rules (Fig. \ref{fig:cycle_schematic}b) suited for hole spin accumulation in a ``dark'' state. We explored the resulting dynamics with a probe beam of identical frequency. For boron acceptors in silicon, the bound hole transitions lie in a frequency range notoriously inaccessible for ordinary laser sources, requiring either a THz pulsed free electron laser or a mixing crystal \cite{Han2025}. Here, we use the FELIX \cite{Oepts1995, Greenland2010} free electron laser to produce $\sim$\SI{9}{\pico\second} coherent pulses in a standard pump-probe geometry (Fig. \ref{fig:cycle_schematic}e) where the pump path length is altered to control the arrival time delay between pump and probe. Polarisation control is achieved via rotating wire-grid polarisers and quarter wave plates made from silicon prisms. 

Fig. \ref{fig:cycle_schematic}a displays the relevant quantum levels. The hole's ground state, $1\Gamma_8^+$, has total angular momentum $J=3/2$, forming a fourfold-degenerate manifold $m_{J}= \pm3/2, \pm1/2$. The excited states in this experiment are $1\Gamma_6^-$ and $1\Gamma_7^-$, both $J=1/2$ doublets. A laser tuned near \SI{9.6}{\tera\hertz} is simultaneously near-resonant with the $1\Gamma_8^+ \to 1\Gamma_6^-$ and $1\Gamma_8^+ \to 1\Gamma_7^-$ transitions as they are separated by only \SI{20}{\giga\hertz}, well within the laser’s $\sim$\SI{40}{\giga\hertz} bandwidth (Fig. \ref{fig:cycle_schematic}b). Both transitions have comparable oscillator strengths \cite{Pajot2010}.

\begin{figure}[b]
    \centering
    \includegraphics[width=1.0\linewidth]{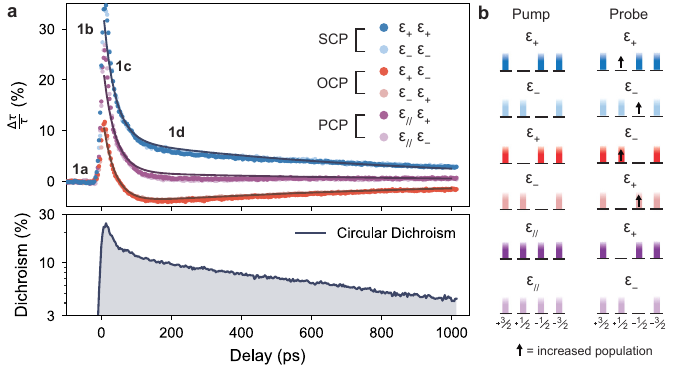}
    \caption{\textbf{Pump-probe raw data and circular dichroism during one initialisation cycle.} \textbf{a} The time-resolved data depicting the change in transmission of the probe $\Delta \tau$ relative to the transmission of a constant reference probe $\tau$. The data come in pairs that overlap almost entirely. The labels \textbf{1a-1d} mark the different parts of the initialisation process in Fig. \ref{fig:cycle_schematic}. At \textbf{1a} all degenerate ground states are in thermal equilibrium and equally populated, representing the transmission baseline. At \textbf{1b} a pump excites a subset of ground states, causing their populations to drop in \SI{9}{\pico\second}. At \textbf{1c} the excited states then quickly decay over the course of $\sim$\SI{36}{\pico\second}. At \textbf{1d} the circularly polarised pumps have created an imbalance in the ground state populations and therefore a dichroism in the probe -- separating red and blue. A linear pump induced no dichroism (purple). Spin-lattice relaxation drives the slow decay back to equilibrium. The sample temperature was \SI{2.9}{\kelvin} and the pump pulse energy was $\sim$\SI{2.9}{\nano\joule}. The model described in the Quantum Dynamics section was fit simultaneously to all the data, and is shown for the SCP, OCP and PCP in the dark solid lines. 
    Below is the circular dichroism, defined as the difference between the SCP (blue) and OCP (red) curves and plotted on a logarithmic scale. \textbf{b} The `bright' ground states excited by the pump and then interrogated by the probe, for all polarisation combinations. An upwards arrow corresponds to that ground state spin population increasing as the excited holes decay after the pump's action.}
    \label{fig:main_data}
\end{figure}

Fig. \ref{fig:main_data} shows the probe's transient transmission change. Three different pump polarisations were used: plane polarised ($\epsilon_{\parallel}$), left circular polarised ($\epsilon_{+}$) and right circular ($\epsilon_{-}$). The probe was either $\epsilon_{+}$ or $\epsilon_{-}$ meaning six combinations in total. The results come in virtually indistinguishable pairs depending on the pump and probe polarisations: same circular polarisation (SCP) for both pump and probe; opposite circular polarisation (OCP); and plane polarised pump, circular probe polarisation (PCP). Each pair produces the same result -- as would be expected for a cubic medium -- and was highly reproducible even though switching polarisation involves a mechanical change of orientation of optics.

Before the pump arrives, all spin populations are at equilibrium and equal (Fig. \ref{fig:cycle_schematic}a). Subsequently, the pump excitation first induces a large transmission increase in the probe as the optically active ground state population (Fig. \ref{fig:cycle_schematic}b, Fig. \ref{fig:main_data}b) is pushed upwards to excited states, resulting in a sudden drop in the density of absorbers and increase in stimulated emission. 

All transients then exhibited fast followed by slow decays. The fast regime emerged as an initial exponential drop in the transmission corresponding to holes rapidly dropping out of the excited orbital (Fig. \ref{fig:cycle_schematic}c) on timescales comparable to those found in previous experiments \cite{Vinh2013,Saeedi2018}. The subsequent slow exponential was either absent (PCP) or decayed towards the baseline either from above (SCP) or below (OCP), indicating that the probe detected a strong circular dichroism induced in the sample by the pump. 

\subsection*{Dark states}

The slow decay component, seen only for circularly polarised pumps and probes, indicates that the dichroism is linked to imbalanced populations of long-lived ground state spins \cite{LUDWIG1962223, Dirksen1989, Stegner2010, Tezuka2010, Song2011, Kobayashi2021}. In systems driven by polarised light, where the ground state manifold is more degenerate than the excited state, as is the situation for B in Si, certain ground-state sublevels may form `dark states' that do not couple to the optical driving field.  
Non-unitary relaxation from the excited state transfers population out of the optically active bright states into the dark spin states. Therefore, when pump and probe have the same polarisation (SCP), they share the same bright and dark states, resulting in reduced probe absorption. With opposite polarisations (OCP), however, the probe’s bright states coincide with the pump’s dark states, yielding enhanced absorption (Fig. \ref{fig:main_data}b). 

The relative transition strengths and dipole moments between states in the $1\Gamma_8^+$ and $1\Gamma_6^-$ or $1\Gamma_7^-$ manifolds have been calculated consistently with group theory \cite{Bhattacharjee1972}. Resolving the spin along $\langle$111$\rangle$, Tables \ref{tab:Qplus} and \ref{tab:Qminus} give the magnitude squared of the electric dipole matrix elements for optical propagation along $\langle$111$\rangle$ for polarisations $\epsilon_{+}$ and $\epsilon_{-}$. 

In contrast to what happens for atoms in vacuum, many of the transitions do not conserve the angular momentum stored in the atom and light subsystem: for instance $m_J=+3/2$ can couple to $m_J=-1/2$ via an $\epsilon_+$ photon. Angular momentum is only conserved if the entire system is considered, including the crystal. Locally, the silicon lattice breaks the rotational symmetry of the bound hole's surroundings, relaxing the conservation laws. The missing angular momentum is transferred to the lattice \cite{Beth1936, bloembergen1980, Tatsumi2018}.

\begin{table}[tbp]
\centering

\small
\renewcommand{\arraystretch}{1.2}
\setlength{\tabcolsep}{4pt}

\newcommand{\mjp}[2]{\raisebox{0.25ex}{$\phantom{-}\tfrac{#1}{#2}$}} 
\newcommand{\mjm}[2]{\raisebox{0.25ex}{$-\tfrac{#1}{#2}$}}   
\newcommand{\z}{\ensuremath{0}}
\newcommand{\sz}{\ensuremath{\mathllap{\sim}\,0}}           

\begin{minipage}[t]{0.47\linewidth}
\centering
{\sffamily
\begin{tabular}{@{}llcccc@{}}
\toprule
\multicolumn{2}{l}{Final state} &
  \multicolumn{4}{l}{\hspace{0.8em}Initial state $1\Gamma_{8}^{+}$} \\ 
\cmidrule(lr){1-2}\cmidrule(l){3-6}
& $m_J$ &
  \hspace{-0.25em}\mjp{3}{2} &
  \hspace{-0.25em}\mjp{1}{2} &
  \hspace{-0.25em}\mjm{1}{2} &
  \hspace{-0.25em}\mjm{3}{2} \\
\midrule
$1\Gamma_6^{-}$ & \mjp{1}{2}   & $\z$           & $\mathbf{0}$ & $\tfrac{1}{4}$ & $\z$ \\
                & \mjm{1}{2}   & $\sz$          & $\mathbf{0}$ & $\z$           & $\tfrac{3}{4}$ \\
\midrule
$1\Gamma_7^{-}$ & \mjp{1}{2}   & $\z$           & $\mathbf{0}$ & $\tfrac{1}{4}$ & $\z$ \\
                & \mjm{1}{2}   & $\tfrac{2}{3}$ & $\mathbf{0}$ & $\z$           & $\tfrac{1}{12}$ \\
\bottomrule
\end{tabular}
}
\caption{$\lvert Q_{+}\rvert^2$ matrix elements along $\langle 111 \rangle$}
\label{tab:Qplus}
\end{minipage}
\hfill
\begin{minipage}[t]{0.47\linewidth}
\centering
{\sffamily
\begin{tabular}{@{}llcccc@{}}
\toprule
\multicolumn{2}{l}{Final state} &
  \multicolumn{4}{l}{\hspace{0.8em}Initial state $1\Gamma_{8}^{+}$} \\
\cmidrule(lr){1-2}\cmidrule(l){3-6}
& $m_J$ &
  \hspace{-0.25em}\mjp{3}{2} &
  \hspace{-0.25em}\mjp{1}{2} &
  \hspace{-0.25em}\mjm{1}{2} &
  \hspace{-0.25em}\mjm{3}{2} \\
\midrule
$1\Gamma_6^{-}$ & \mjp{1}{2}   & $\tfrac{3}{4}$   & $\z$           & $\mathbf{0}$ & $\sz$ \\
                & \mjm{1}{2}   & $\z$             & $\tfrac{1}{4}$ & $\mathbf{0}$ & $\z$ \\
\midrule
$1\Gamma_7^{-}$ & \mjp{1}{2}   & $\tfrac{1}{12}$  & $\z$           & $\mathbf{0}$ & $\tfrac{2}{3}$ \\
                & \mjm{1}{2}   & $\z$             & $\tfrac{1}{4}$ & $\mathbf{0}$ & $\z$ \\
\bottomrule
\end{tabular}
}
\caption{$\lvert Q_{-}\rvert^2$ matrix elements along $\langle 111 \rangle$}
\label{tab:Qminus}
\end{minipage}

\end{table}

We see in Table \ref{tab:Qplus} that there is only one component of the $1\Gamma_{8}^{+}$ ground quartet that is fully dark when pumping $1\Gamma_{6}^{-}$ and $1\Gamma_{7}^{-}$ simultaneously, which is  $\ket{1\Gamma_{8}^{+},m_J=1/2}$. It can be shown that there exist no other $m_{J}$ spin states, nor possible superpositions of them, which are dark (See Supplementary Note 1). All of the zeros in Table \ref{tab:Qplus} are exact by symmetry except for $\braket{\Gamma_{6}^{-},m_J=-1/2|Q_{+}|\Gamma_{8}^{+},m_J=3/2}$, which is proportional to a dimensionless Zeeman mixing parameter that is small. The conclusion that $\ket{1\Gamma_{8}^{+},m_J=1/2}$ is the only dark state when pumping $1\Gamma_{6}^{-}+1\Gamma_{7}^{-}$ is guaranteed by symmetry and is independent of any parameters.

In our experiment, optical pumping with $\epsilon_{+}$-polarised photons propagating along $\langle 111 \rangle$ transfers population efficiently into the \(\ket{1\Gamma_{8}^{+},m_J=1/2}\) component, while $\epsilon_{-}$ transfers population into \(\ket{1\Gamma_{8}^{+},m_J=-1/2}\). This outcome is especially noteworthy because it represents the only mechanism for preparing a unique dark state in \emph{unstrained} silicon. In contrast, pumping via any other state (Fig. \ref{fig:independence}, Supplementary Fig. S4) along any direction or pumping the $1\Gamma_{6}^{-}+1\Gamma_{7}^{-}$ manifold along the $\langle100\rangle$ or $\langle110\rangle$ axes can achieve, at best, partial spin polarisation, since in those geometries there are either no dark states or multiple dark states, preventing selective spin preparation.

\subsection*{Orbital relaxation}
 
Excited holes decay rapidly back to the degenerate ground states: below \SI{10}{\kelvin} both the $1\Gamma_{6}^{-}$ and $1\Gamma_{7}^{-}$ states have large natural line-widths and fast relaxations on the order of $\sim$\SI{36}{\pico\second} \cite{Vinh2013, Saeedi2018}. This decay can happen via photon emission just as for free atoms, but phonon-mediated transitions dominate impurity relaxation \cite{Lax1960, Ridley2013}. Similarly with photon transitions, missing angular momentum is transferred to the lattice \cite{Zhang2014, Garanin2015, Ruckriegel2020}. 

The details of such processes are unclear, since there are no phonons resonant with $1\Gamma_8^+\leftrightarrow 1\Gamma_6^-+1\Gamma_7^-$ that conserve crystal momentum. Therefore it is typically assumed \cite{Lax1960, Pavlov2000, Vinh2013, Saeedi2018} that the relaxation follows a multi-phonon cascade via several intermediate states, the exact details of which are still unknown. Furthermore, the selection rules for phonon transitions are much less stringent than for our optical pumping, as the emitted phonons have no fixed polarisation or direction. As such, for the purposes of modelling we will assume that the orbital decay is a random, uniform decay from all excited states to all ground states.

\subsection*{Spin relaxation measured with THz circular dichroism}

Following optical excitation and orbital decay, there is now a larger population in the target dark state, responsible for the long-lived dichroism in Fig. \ref{fig:main_data}. 
Spin-lattice relaxation (Fig. \ref{fig:cycle_schematic}d) is responsible for the decoherence and equilibration of the ground state spins over times of order \SI{1}{\nano\second}, thus recovering the signal baseline. We also investigated dependence on optical power: varying the pump intensity over an order of magnitude produces negligible change in the long-lived dichroism component (Fig. \ref{fig:simulations_extrapolated}), in contrast to the strong power dependence seen for conventional linearly polarised pumping, thought to arise from long-lived states in the continuum \cite{Vinh2013}. Pumping directly to the continuum was never observed to induce circular dichroism, confirming that dichroism arises solely for intra-acceptor processes (see Supplementary Note 2).

\begin{figure}[t]
\centering
\includegraphics[width=1.0\linewidth]{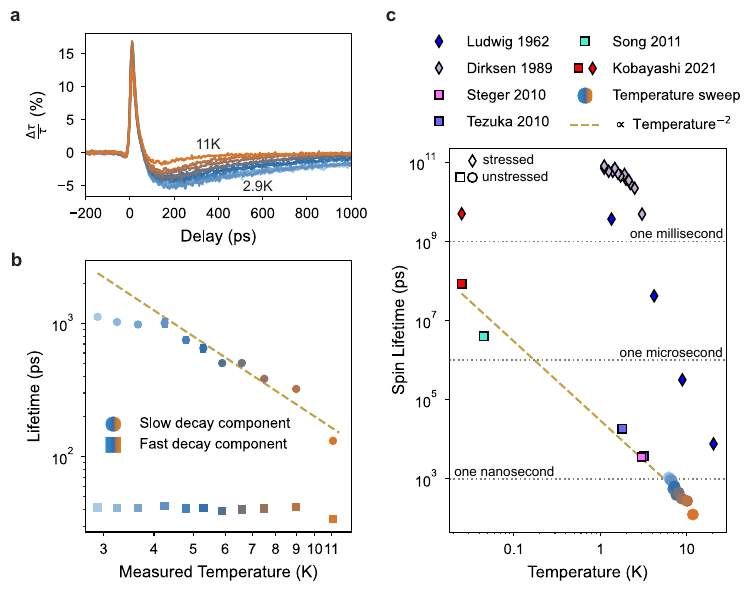}
\caption{
\textbf{Temperature dependence of spin and orbital lifetimes.} 
\textbf{a} OCP data for various temperatures between \SI{2.9}{\kelvin} and \SI{11}{\kelvin}. \textbf{b} Fast (orbital) and slow (spin) lifetimes as a function of temperature, extracted from the OCP data in \textbf{a}. Error bars represent the standard deviations, and are mostly smaller than the symbols. \textbf{c} Spin lifetimes from \textbf{b} are compared to those from previous microwave experiments ($T_{1}$, extracted either from time-resolved spin-echo measurements or from CW EPR linewidths, see Methods): Ludwig 1962 (stress: 490$\pm$\SI{30}{\mega\pascal}) \cite{LUDWIG1962223}, Dirksen 1989 (stress: \SI{750}{\mega\pascal}) \cite{Dirksen1989}, Stegner 2010 \cite{Stegner2010}, Tezuka 2010 \cite{Tezuka2010}, Song 2011 \cite{Song2011} and Kobayashi 2021 (both unstressed and \SI{30}{\mega\pascal} stress) \cite{Kobayashi2021}. 
Our experimental data are represented by circles. Literature results for unstrained silicon are shown as squares and for strained silicon as diamonds. Here we have applied a small correction to the measured temperatures for our data points (denoted by circles). The correction grows at lower temperatures as the sample temperature (exposed to background radiation through the cryostat windows) saturates while the (unexposed) thermometer temperature continues to drop. Details are in Supplementary Note 4.
}

\label{fig:temp_dependence}
\end{figure}

In theory the spins in the ground-state manifold are degenerate. However, inhomogeneous broadening in the crystal splits the ground state quartet \cite{Stegner2010, Tezuka2010}, permitting single- and multi-phonon decay. Fig. \ref{fig:temp_dependence}c compares our temperature-dependent spin-lattice relaxation times to those from previous boron electron paramagnetic resonance (EPR) measurements for both unstrained (the case for our experiments) and strained silicon hosts. Our pump-probe technique allows us to establish time-resolved spin-lattice relaxation using optical THz pulses which can measure dynamics in the sub-10 nanosecond range not reachable with traditional pulsed microwave EPR. 
A fit with the spin relaxation rate rising quadratically with temperature is also plotted alongside all the measured lifetimes, suggesting a two-phonon Raman process \cite{Shrivastava1983, Singh1979}.

\begin{figure}[htp]
\centering
\includegraphics[width=0.9\linewidth]{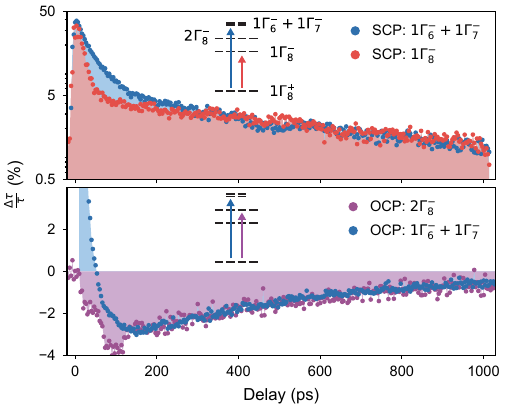}
\caption{
\textbf{Optical pumping through different states.} \textit{Top}: SCP data (logarithmic y-axis) for transitions into the $1\Gamma_{6}^{-} + 1\Gamma_{7}^{-}$ states (blue) and into the $1\Gamma_{8}^{-}$ excited state (red). The fast initial decay was dependent on the excited state, but the final slow decay was not and closely matched to within each other's standard deviation. The pump energy for $1\Gamma_{6}^{-} + 1\Gamma_{7}^{-}$ was \SI{1.6}{\nano\joule}, for $1\Gamma_{8}^{-}$ it was \SI{5.9}{\nano\joule}, reflecting the difference in oscillator strengths \cite{Pajot2010}. \textit{Bottom}: OCP data for $2\Gamma_{8}^{-}$ compared to normalised OCP data for $1\Gamma_{6}^{-} + 1\Gamma_{7}^{-}$. The long lifetime components also match. 
}

\label{fig:independence}
\end{figure}

We also investigated optical pumping via other excited states, such as $1\Gamma_8^-$ and $2\Gamma_8^-$ \cite{Pajot2010}, at \SI{7.3}{\tera\hertz} and \SI{8.3}{\tera\hertz} above the ground state, respectively. In unstrained silicon, only the $1\Gamma_6^-+1\Gamma_7^-$ pathway allows for a single dark state. However, pumping via $1\Gamma_8^-$ or $2\Gamma_{8}^{-}$ also leads to partial spin polarisation and features long spin transients (Fig. \ref{fig:independence}). The long transient had the same decay rate for $1\Gamma_8^-$, $2\Gamma_8^-$ and $1\Gamma_6^-+1\Gamma_7^-$ excitations, despite their distinct selection rules, indicating that spin equilibration within the ground-state manifold is largely insensitive to the details of the population distribution. In our modelling, we therefore treat spin relaxation as a stochastic process characterised by a single effective lifetime. Interestingly, $2\Gamma_8^-$ was the only state with \emph{no} positive component for the OCP measurement, suggesting that there must be two dark ground states, which is not clear from symmetry arguments alone \cite{Bhattacharjee1972}. 

\subsection*{Quantum Dynamics Model}

To extract characteristic lifetimes of this hydrogenic system, and to predict the time-scales of full state initialisation, we modelled the dynamics using a Lindblad master-equation. By taking calculated optical-selection rules, and assuming uniform relaxation of orbital and spin degrees of freedom, we reduced the problem to four free parameters: the Rabi frequencies of the $1\Gamma_8^+\rightarrow1\Gamma_6^-$ and  $1\Gamma_8^+\rightarrow1\Gamma_7^-$ transitions, the orbital decay rate from the excited to the ground states, and the spin-relaxation rate between ground-state sublevels. We performed a fit to the entire dataset in Fig. \ref{fig:main_data}a using this shared set of parameters, and then to each OCP scan in Fig. \ref{fig:temp_dependence}a to extract the relaxation time dependences on temperature, shown in Fig. \ref{fig:temp_dependence}b. Further details are provided in Supplementary Note 1. 

\begin{figure}[htp]
\centering
\includegraphics[width =1.0\linewidth]{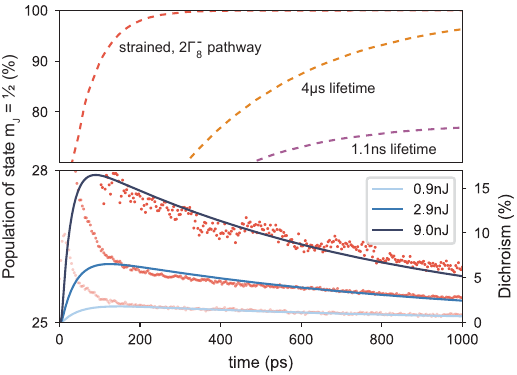}
\caption{
\textbf{Fits and simulations of target population.} \textit{Bottom}: The blue solid lines represent the quantum dynamics model result for the population of the target state $m_{J}=1/2$ for three different pump pulse energies: \SI{9.0}{\nano\joule}, \SI{2.9}{\nano\joule} and \SI{0.9}{\nano\joule}, over the course of \SI{1}{\nano\second} from the moment the pump arrives. The dichroism data for each energy are plotted alongside the modelled populations in red. We observed little power dependence of the spin relaxation time $T_{1}$ measured by our fits: the \SI{9.0}{\nano\joule}, \SI{2.9}{\nano\joule} and \SI{0.9}{\nano\joule} pumps displayed $T_{1}=1020\pm19$, $1136\pm15$ and $1129\pm15$ picoseconds respectively. 
\textit{Top}: The dashed lines are simulations of the target state population throughout a hypothetical \SI{1}{\nano\second} laser pulse. The simulation in purple used the same parameters extracted from the \SI{2.9}{\nano\joule} pump, apart from the pulse duration.  In orange the spin lifetime was replaced with the value from a very similar boron-doped silicon sample at \SI{45}{\milli\kelvin}  \cite{Song2011}. In red is our simulation for an optimised scheme using strain and the $2\Gamma_{8}^{-}$ state as the intermediate state for optical pumping (see Supplementary Note 3).
}
\label{fig:simulations_extrapolated}
\end{figure}

In Fig. \ref{fig:main_data}a, the four-parameter model is displayed alongside the data, and accurately captures the behaviour and lifetimes. The long spin lifetime was $T_{1}=1136\pm15$\SI{}{\pico\second}, while the short orbital relaxation lifetime was 36.1$\pm$0.6\SI{}{\pico\second}. The ratio between the transition rates to $1\Gamma_{6}^{-}$ and $1\Gamma_{7}^{-}$ was 0.104$\pm$0.003, as our laser's central frequency was set slightly below the lower of the two transitions, resulting in stronger coupling to $1\Gamma_{6}^{-}$. Using fitting parameters extracted from the data of several pump energies, we simulated the time evolution of the hole population in the target dark state (Fig. \ref{fig:simulations_extrapolated}). At the highest pump energy (\SI{9.0}{\nano\joule}), we estimated that the dark state population increased from 25\% at equilibrium to 27.9\% just after one pump cycle, corresponding to a roughly 12\% increase in target spin population.

We then extended the simulations to model the target spin population throughout a \SI{1}{\nano\second} pulse and estimate the pulse duration required for 99\% initialisation. Applying the fitting parameters extracted from the intermediate $\sim$\SI{2.9}{\nano\joule} pump energy, and extending the pulse duration to \SI{1}{\nano\second} while maintaining pump power ($\sim$\SI{0.3}{\kilo\watt}), the population saturates at 77\% within \SI{1}{\nano\second}. The achievable fidelity reflects the competition between optical pumping and spin relaxation processes: no matter the pumping power or duration, the initialised population never exceeds 78\% unless the spin lifetime increases beyond our measured $T_{1}=1136\pm15$\SI{}{\pico\second}. We repeated the simulation using a spin lifetime of \SI{4}{\micro\second} -- the $T_{1}$ lifetime measured at \SI{45}{\milli\kelvin} in a very similar boron-doped silicon sample \cite{Song2011}. In this case, the initialised fraction surpasses 99\% within \SI{1.4}{\nano\second}. Both scenarios are presented in Fig. \ref{fig:simulations_extrapolated}. Our simulations do not account for the slightly imperfect beam circularity, so in fact present a lower bound for the rate at which states can be initialised.

Under strain, all $\Gamma_{8}$ quartets split into two Kramer's doublets $\pm3/2$ and $\pm1/2$. This opens up many new pathways for optical pumping, including the short-lived $2\Gamma_{8}^{-}$ (orbital $T_{1} \approx 15.0\pm0.8$\SI{}{\pico\second}, see Supplementary Fig. S4) with a very large oscillator strength. Furthermore, the new ground state -- an isolated $\pm1/2$ -- has been shown to exhibit considerably improved coherence times, exceeding \SI{10}{\milli\second} \cite{Salfi2016, Song2011, Ruskov2013, Kobayashi2021}. There is also evidence for $>$ \SI{10}{\milli\second} $T_{1}$ lifetimes at high temperatures ($>$ \SI{1}{\kelvin}) and high stress, shown in Fig. \ref{fig:temp_dependence}c \cite{Dirksen1989, LUDWIG1962223}. With these improvements we calculate that it would take under \SI{250}{\pico\second} to exceed 99\% initialisation (Fig. \ref{fig:simulations_extrapolated}) at $\sim$\SI{3}{\kelvin}. Evidence of a much shorter $T_{1}=$ \SI{6.3}{\pico\second} \cite{Saeedi2018} would result in a spin initialisation time of $\sim$\SI{124}{\pico\second}. More details are given in Supplementary Note 3. 

\section*{Discussion and Outlook}

We have demonstrated that strong spin-orbit coupling in silicon's valence band can be harnessed to rapidly initialise and read out the spin states of acceptor-bound holes using optical pumping of low hydrogen-like Rydberg states. This was performed on a standard boron-doped silicon sample, and required THz circular dichroism to probe spins in solid state systems. 

Using a remarkably simple atomic model with only four free fitting parameters, we could simultaneously fit to several configurations of pump-probe transmission data and extract estimates of population evolution \SI{1}{\nano\second} after pumping. We measured power and temperature dependences to explore the capabilities and limitations of this method. The dark state population increased by 12\% after a single \SI{9}{\pico\second} pump and $\sim$\SI{36}{\pico\second} orbital decay, leading us to expect high-fidelity spin initialisation to work on sub-nanosecond timescales. 

Looking ahead, a key milestone would be to combine rapid optical spin manipulation with the much longer spin lifetimes observed when the ground states are split by strain. We would then expect near-total initialisation to be reached in under \SI{250}{\pico\second} at temperatures above \SI{3}{\kelvin}. 
This is at least three orders of magnitude faster than the shortest electrical initialisation times measured in silicon-based quantum dots or dopants which still suffer from strict cryogenic requirements and scaling difficulties \cite{Blumoff2022, Stano2022, Burkard2023, Gonzalez-Zalba2021, Reiner2024}.
This is likely to improve further with acceptors in silicon with even shorter orbital lifetimes -- such as indium or aluminium -- or in acceptors in semiconductors with stronger spin-orbit coupling such as germanium \cite{Ramdas2001}. These have the added advantage of orbital transition frequencies reachable by commercially available tabletop laser sources. 

To our knowledge optical pumping has not been attempted for acceptor centres in any elemental semiconductor. Partial polarisation has been achieved in silicon \emph{donors} using an indirect, ionising, and much slower $\sim$\SI{100}{\milli\second} process \cite{Yang2009, Lo2015}. Initialising other defects in diamond, SiC and GaAs is also much slower \cite{Pingault2017, Robledo2011, Green2017, Linpeng2021, Nagy2019}. Optical initialisation of single hole spins in III–V quantum dot heterostructures, enabled by spin-selective excitonic transitions, has been extensively demonstrated \cite{Gerardot2008, Greilich2011, Dusanowski2022, Brunner2009}. Trapped-ion and neutral atom platforms both require motional cooling (millisecond timescales \cite{Eschner2003Laser}) and state initialisation before computation \cite{Bruzewicz2019Trapped}, typically performed by optical pumping over tens of microseconds.

By accessing the orbital degree of freedom, we increase the natural energy scales from the GHz to the THz regime, and the natural time scales from nano- to picoseconds. This relaxes thermal constraints and both initialisation as well as readout become effectively instantaneous as compared to current semiconductor quantum gate and measurement times. Here we have focused on initialisation and readout via circular dichroism, but single and two-qubit operations for single acceptors clearly warrant future research. These could build on mixed microwave-optical or all-optical paradigms already proposed for donors in silicon \cite{Stoneham2003, Crane2021} and other single atom-based systems \cite{Grimm2021}, but now with the added benefits of large spin-orbit coupling. The fact that there is a deterministic process to substitute individual boron for silicon atoms \cite{Skeren2020}, analogous to that well established for phosphorous \cite{Shen1995, Wilson2004, Wyrick2019} 
is an advantage for this defect over others, such as T-centres \cite{Higginbottom2022} and interstitial rare earths \cite{Yin2013}, which have optical transitions of potentially similar utility but at more common wavelengths. Also, in our experiments, we performed an ensemble average where the laser's global coverage prepared and measured all spins simultaneously, but suitable structures including both patterned on-chip metals \cite{Akhlaghi2015} as well as single electron transistors acting as local detectors \cite{Zwanenburg2013} could enable single spin addressability.

\section*{Methods}

The spin preparation and time-resolved measurements were all performed within a standard pump-probe set-up \cite{Greenland2010, Vinh2013}. Each $\sim$\SI{9}{\pico\second} FELIX pulse was split into a pump and probe channel. The pump and probe could be independently set to $\epsilon_{+}$, $\epsilon_{-}$ and $\epsilon_{\parallel}$ polarisations. This was achieved with a wire-grid polariser on a motorised rotator followed by a Fresnel retarder made from a right-angle silicon prism behaving as a quarter wave plate. This silicon prism outputs linearly polarised light for a total internal s-- or p-- reflection; however, for a $\pm45\degree$ input polarisation, the output is nearly $\epsilon_{\pm}$ circularly polarised \cite{Stanislavchuk2013} (our photon polarisation circularities for $\epsilon_{+}$ and $\epsilon_{-}$ were 84\% and 90\% respectively in electric field amplitude). A delay stage and set of attenuators were included for time-resolution and power control. To measure the power before the cryostat we used a calibrated pyroelectric sensor (SLT PEM 34 IR); to calculate the pulse energy inside the sample we corrected this value by accounting for the absorption of the cryostat windows and the reflection from the silicon surface. 

Both beams overlapped on the sample surface with $\sim$\SI{1}{\milli\meter} Gaussian profiles. This was ensured with a pyroelectric camera. They were aligned to the $\langle111\rangle$ crystal axis, with slight deviations $<5\degree$ to block any pump beam from leaking into the probe's detector channel. The sample was an \SI{800}{\micro\meter} thick, $3\times10^{15}$cm$^{-3}$ boron-doped, FZ--silicon sample and was mounted with silver paste onto a \SI{3}{\kelvin} cold finger in vacuum, measured by a sensor mounted on the same cold finger. It was cut with its normal parallel to $\langle 111 \rangle$. The data in the top panel of Fig. \ref{fig:independence} were collected using an isotopically purified $^{28}$Si sample, thickness \SI{430}{\micro\meter}, with boron doping concentration $2.4\times10^{15}$ cm$^{-3}$, cut with its normal parallel to $\langle100\rangle$. 

The $T_{1}$ lifetimes displayed in Fig. \ref{fig:temp_dependence}c were either lifted directly from the literature \cite{LUDWIG1962223, Dirksen1989, Song2011, Kobayashi2021} or were lower bound estimates of $T_{1}$ \cite{Stegner2010, Tezuka2010}, using the relation $T_{1} \geq T_{2}/2 \geq T_{2}^{*}/2$. Estimates for $T_{2}$ were obtained from the peak-to-peak linewidths of spectra measured via CW EPR of isotopically purified silicon samples, for which there was little inhomogeneous broadening. 

\section*{Author Contributions}
A.G.M., G.M. and G.A. conceived the project. S.G.P. and N.V.A. provided and characterised samples. N.D., A.G.M., S.G.P. and G.M. prepared and carried out the experiments with input from G.A. and L.A. A.G.M. and N.D. analysed the data. A.G.M. and W.A. worked on the quantum dynamics simulations with input from L.A. and G.A. Theoretical understanding was developed by A.G.M., B.N.M., W.A. and L.A. A.G.M., W.A., G.A., B.N.M., N.D. and G.M. wrote the manuscript with input from all co-authors.

\section*{Acknowledgements}
A.G.M. and G.A. acknowledge funding for the current project from the European Research Council under the European Union’s Horizon 2020 research and innovation program, within Grant Agreement 810451 (HERO). L.A. was funded by the Swiss National Science Foundation under Grants No. $200020\_200558$ and No. $200021\_166271$. B.N.M acknowledges support from EPSRC-UK Grant number EP/M009564/1 and the European Research Council under the European Union’s Horizon 2020 research and innovation program, within Grant Agreement 810451 (HERO). 



\bibliography{bibliography}

\end{document}



\begin{center}
\LARGE \textbf{Supplementary Information}\\[4pt]
\large Rapid high-temperature initialisation and readout of spins in silicon with 10 THz photons\\
\end{center}

\beginsupplement


\section*{Supplementary Note 1: Quantum Dynamics Model}

Any quantum computation can be divided into three steps: state preparation, computation and detection. Here we focus on state preparation: a process which takes an initially mixed quantum state and prepares it in a pure quantum state through dissipative entropy extraction. We model our system -- a hole bound to a boron atom embedded in a silicon matrix -- as an atom in free space with spin-orbit (SO) coupling, replacing only the vacuum-photon radiative decay with non-radiative phonon decay. The selection rules are also adjusted to account for the broken symmetry due to the crystal. 

\subsection*{Free-Space Atom Optical Pumping}

We borrow the concept of dark state optical pumping from atomic physics. To prepare the spin of an atom in vacuum, specific atomic transitions are selectively driven by light of frequency $\omega$ and polarisation $\epsilon$. In contrast, spontaneous relaxation is mediated by vacuum fluctuations containing all polarisations and frequencies, and is therefore much less constrained by selection rules. This asymmetry can be exploited to prepare a well-defined pure state of the atom. 

When one of the ground states is a dark state of the light-matter interaction Hamiltonian, but is still accessible through spontaneous decay from the excited manifold, the population gradually accumulates there. For instance, in a system with $J=3/2$ in the ground state and excited state manifolds $g$ and $e$, illumination with $\epsilon_{+}$ polarised light leaves $\ket{g, J=\frac{3}{2}, m_J=\frac{3}{2}}$ as the only dark state. Over time, repeated excitation and decay funnels population here.

Applying a magnetic field provides additional control by shifting unwanted transitions off resonance, further enhancing state selectivity beyond polarisation control. In this work, we realise similar dark-state pumping in boron atoms embedded in silicon, without the application of an external magnetic field.

\subsection*{Selection rules and dark states}

Without splitting the ground state quartet with magnetic/electric fields or strain, the only way to isolate a single dark state in acceptors in silicon for optical pumping is by coupling the $1\Gamma_{8}^{+}$ ground state quartet to both the $1\Gamma_{6}^{-}$ and $1\Gamma_{7}^{-}$ doublets simultaneously with circularly polarised light propagating along a $\langle 111 \rangle$ axis. This is possible because these doublets are separated by a smaller energy gap than our pulsed laser's spectral width, and both have similar oscillator strengths \cite{Pajot2010}. Linear polarisation, other propagation directions or pumping via any other excited state would result in either two or zero dark states. 

A \emph{dark state} is a coherent superposition of states in the ground level whose electric–dipole coupling to applied light field vanishes, so the state neither absorbs nor emits light. A four-level ground manifold and an excited doublet that are coupled optically can, through the Morris–Shore transformation \cite{Morris1983}, each be decomposed into an orthonormal set with only two allowed transitions. The ground level, therefore, comprises two `bright states' and two dark states. In the present system, both $1\Gamma_{8}^{+}\to1\Gamma_{6}^{-}$ and $1\Gamma_{8}^{+}\to1\Gamma_{7}^{-}$ produce a pair of dark states. For spin initialisation, however, it is desirable to isolate a single dark state. This can, in general, be achieved by lifting the degeneracy of either the ground or excited levels using a magnetic or electric field, or via strain. In our case such external perturbations are unnecessary: the two transitions share only one dark state in common (which is a pure basis state) $\ket{1\Gamma_{8}^{+}, m_J=1/2}$ in the case of $\epsilon_+$ (LCP) light while the remaining dark state differs for each transition, leaving a single global dark state in the ground manifold when both transitions are pumped simultaneously. In the case of $\epsilon_-$ (RCP) light the global dark state is $\ket{1\Gamma_{8}^{+}, m_J=-1/2}$.

We first consider $\Gamma_8\rightarrow\Gamma_6$ transitions for circularly polarised light propagating along  $[111]$.
From \cite{Bhattacharjee1972} we have the dipole moment matrices (valid at small and zero fields) $\braket{e|Q_x|g}$ and $\braket{e|Q_y|g}$ for $x-$ and $y-$polarised light propagating along [111], defined to be the $z$ axis. 
Here the ground states $\ket{g} \in \{ \ket{\Gamma_8^+,m_J}\}$ have $m_J\in\{3/2,1/2,-1/2,-3/2\}$ running along the columns from left to right, and the excited states $\bra{e} \in \{ \bra{\Gamma_6^-,m_J}\}$ have $m_J\in\{1/2,-1/2\}$ running down the rows from top to bottom (as in the main text tables 1 \& 2):
\[
Q_{x}(\Gamma_{8\rightarrow6})=D_0\begin{bmatrix}
\sqrt{3}\,q & 0 & 1 & i\sqrt{3}\,\gamma q\\
-\sqrt{3}\,\gamma q & i & 0 & i\sqrt{3}\,q
\end{bmatrix},\qquad
Q_{y}(\Gamma_{8\rightarrow6})=D_0\begin{bmatrix}
i\sqrt{3}\,q & 0 & -i & -\sqrt{3}\,\gamma q\\
i\sqrt{3}\,\gamma q & -1 & 0 & \sqrt{3}\,q
\end{bmatrix},
\]
where $q=1/\sqrt{1+\gamma^2}$ and $D_0$ is an orbital dipole moment integral which gives an overall scale for the transition rate. From these we form the circularly polarised dipole moment matrices
\[
\begin{aligned}
    Q_{+}(\Gamma_{8\rightarrow6}) &=\frac{Q_{x}+iQ_{y}}{\sqrt{2}} = D_0\begin{bmatrix}0&0&\sqrt{2}&0\\ -\sqrt{6}\,\gamma q & 0&0 & i\sqrt{6}\,q\end{bmatrix},\\
Q_{-}(\Gamma_{8\rightarrow6}) &=\frac{Q_{x}-iQ_{y}}{\sqrt{2}}= D_0\begin{bmatrix}\sqrt{6}\,q&0&0&i\sqrt{6}\,\gamma q\\ 0&i\sqrt{2}&0&0\end{bmatrix},
\end{aligned}
\]
The matrix elements are written in terms of a parameter $\gamma=\sqrt{2} \left(\beta-\sqrt{\beta^2+\frac{1}{2}}\right)$, which determines the mixing of the $m_J=\pm3/2$ components of the $1\Gamma_8^+$ ground state, where $\beta=\frac{3g_1'}{2 g_2'}+\frac{23}{8}$. $g_1'$ and $g_{2}'$ are the isotropic and anisotropic parts respectively of the g-factor for the ground state and are reported by \cite{Kopf1992},  yielding $\gamma \sim -0.0069$. For simplicity in the main text we assume $\gamma\approx 0$ and $q\approx1$. The normalised square magnitude of the elements $Q_{\pm}$ give the relative intensities stated in \cite{Bhattacharjee1972} and reproduced in tables 1 \& 2 in the main manuscript.

Following the same procedure to obtain the circular dipole moments for $\Gamma_8\rightarrow\Gamma_7$ transitions where $\bra{e} \in \{ \bra{\Gamma_7^-,m_J}\}$ requires correction of a typographical sign error in \cite{Bhattacharjee1972}, which we presume to be in the $Q_{x}$ matrix, giving:
\[
Q_{+}(\Gamma_{8\rightarrow7})
=D_0'
\begin{bmatrix}
0&0&i\sqrt{2}&0\\
-i\,\sqrt{\tfrac{2}{3}}(2\sqrt{2}+\gamma)q&0&0&\sqrt{\tfrac{2}{3}}(2\sqrt{2}\gamma-1)q
\end{bmatrix},
\]
\[
Q_{-}(\Gamma_{8\rightarrow7})
=D_0'
\begin{bmatrix}
i\sqrt{\tfrac{2}{3}}(2\sqrt{2}\gamma-1)q&0&0&\sqrt{\tfrac{2}{3}}(2\sqrt{2}+\gamma)q\\
0&-\sqrt{2}&0&0
\end{bmatrix}.
\]
In what follows we shall make use of the definition 
\[ \alpha = \frac{D_0'}{D_0}\]
so that $D_0'=\alpha D_0$.

We see that for $\epsilon_{+}$ light the dipole moment matrix element $\braket{e|Q_+|g}$ for $\ket{g}=\ket{1\Gamma_8^+,1/2}$ is zero for \emph{all} possible excited states, by symmetry. This ground state is trivially dark. We also see that $\ket{g}=\ket{1\Gamma_8^+,-1/2}$ is bright and can be excited to both $\bra{e}=\bra{1\Gamma_{6,7}^{-},1/2}$ excited states. Finally, we see that there is a superposition $\ket{g}=a\ket{1\Gamma_8^+,3/2}+b\ket{1\Gamma_8^+,-3/2}$ that can be constructed which also yields a transition rate of zero via destructive interference. In the case of $\Gamma_6^-$ we find $\binom{a}{b}=q\binom{1}{-i\gamma}$ is dark and $q\binom{-i\gamma}{1}$ is bright (and orthogonal to the dark state). In the case of $\Gamma_7^-$ we find $\binom{a}{b}=\tfrac q3\binom{1-\sqrt{8}\gamma}{-i(\sqrt{8}+\gamma)}$ is dark and $\binom{a}{b}=\tfrac q3\binom{-i(\sqrt{8}+\gamma)}{1-\sqrt{8}\gamma}$ is bright. There is no value of $\gamma$ that makes the two dark superpositions coincident (the inner product of the superposition bright state for one excited level with the superposition dark state for the other has magnitude $\sqrt{8}/3$ and is independent of $\gamma$). This symmetry constraint leaves $\ket{g}=\ket{1\Gamma_8^+,1/2}$ as the only dark state for $\epsilon_{+}$ light and $\ket{g}=\ket{1\Gamma_8^+,-1/2}$ as the dark state for $\epsilon_{-}$.

\begin{figure}[htp]
    \centering
    \includegraphics[width=1\linewidth]{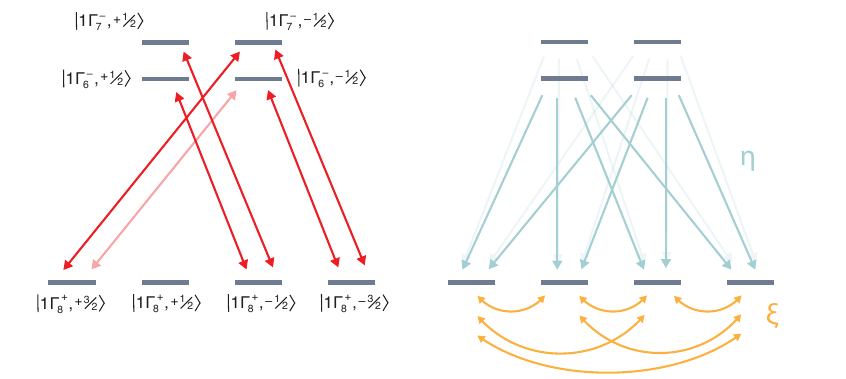}
    \caption{\textbf{Selection rules for photons and phonons.} Sketch of the energy levels with indicated couplings and selection rules for $\epsilon_{+}$ photons (left) and unpolarised phonons (right), using the assumptions made in the text.
    }
    \label{fig:8_level_sketch}
\end{figure}

The selection rules are illustrated in Fig. \ref{fig:8_level_sketch}. They are altered by broken symmetry, for example by introduction of strain and magnetic field, but still allow for optical pumping. Introducing such perturbations would allow for more pumping opportunities and schemes, as highlighted in Supplementary Note 3 below.

\subsection*{Model}

The spin-initialisation process is modelled using a Lindblad master equation, reflecting the intrinsically non-unitary nature of entropy reduction during optical pumping
    \begin{equation}
    \frac{d \rho}{d t}=\frac{1}{i \hbar}[H, \rho]+\mathcal{L}(\rho)
    \label{lind_equation}
\end{equation}
where the Hamiltonian $H$ can be decomposed into:
\[H=H_o+H_d+H_B\]
We take the rotating wave approximation and work in the interaction picture rotating with the laser drive. The off-diagonal term $H_o$ describes coherent optical couplings:
\[H_o = \sum_{g\in \{1\Gamma_{8}^{+}\}, \,e\in \{1\Gamma_{6}^{-},\,1\Gamma_{7}^{-}\}} 
\frac{1}{2}\hbar
\left(\Omega_{eg}\ket{e}\bra{g}+\Omega_{eg}^*\ket{g}\bra{e}\right).\]
The Rabi frequencies are $\Omega_{eg}=Q_{eg}E/\hbar$ where $E$ is the electric field amplitude of the pump and $Q_{eg}$ is, as above, the transition dipole moment between states $e \text{ and } g$. We define $\Omega_0=D_0E/\hbar$ so that $\Omega_{eg}=\Omega_0  Q_{eg}/D_0$ where dimensionless dipole moments $Q_{eg}/D_0$ are given in the previous section in terms of the parameters $\gamma$ and $\alpha$. 
The diagonal term 
\[H_d = \sum_{e\in \{1\Gamma_{6}^{-},\,1\Gamma_{7}^{-}\}} \hbar\Delta_{e} \ket{e}\bra{e}\]
accounts for the detuning of the laser beam with respect to the energy separation of the excited states from the ground states. The last term $H_B = \sum_{i} g_{J}^{(i)} \mu_{B} m_J^{(i)} B \ket{i}\bra{i},$ represents Zeeman shifts. In the simulation presented here we take $B=0$. 

Dissipation is described by the Lindblad term

\[
\begin{aligned}
\mathcal{L}(\rho)
&= \underbrace{- \frac{1}{2}\sum_{g\in \{1\Gamma_{8}^{+}\}, \,e\in \{1\Gamma_{6}^{-},\,1\Gamma_{7}^{-}\}}\left(( c^{(\eta)}_{eg})^\dagger c^{(\eta)}_{eg} \rho + \rho (c^{(\eta)}_{eg})^\dagger c^{(\eta)}_{eg} -2 c^{(\eta)}_{eg}\rho (c^{(\eta)}_{eg})^\dagger\right)}_{\text{Orbital Decay}}\\
&\quad \underbrace{- \frac{1}{2}\sum_{g, g'\in \{1\Gamma_{8}^{+}\}}\left(( c^{(\xi)}_{gg'})^\dagger c^{(\xi)}_{gg'} \rho + \rho (c^{(\xi)}_{gg'})^\dagger c^{(\xi)}_{gg'} -2 c^{(\xi)}_{gg'}\rho (c^{(\xi)}_{gg'})^\dagger\right)}_{\text{Spin Relaxation}} \, ,
\end{aligned}
\]

where phonon-mediated relaxation and ground state mixing are defined as represented respectively by the jump operators  $c^{(\eta)}_{eg} = \sqrt{\eta}\ket{g}\bra{e}$ and $c^{(\xi)}_{gg'} = \sqrt{\xi}\ket{g'}\bra{g}$. 
The two dissipation processes included are the orbital relaxation from excited state to ground state, which we assume all have the same rate $\eta_{eg}=\eta$, and the spin relaxation among the ground states, again assumed all equal $\xi_{gg'}=\xi$, as shown in Fig. \ref{fig:8_level_sketch} and explained in the main text.

Time-dependent simulations of the eight-level system were performed using `QuTiP'. The system was initialised as an equal mixture of all ground states, and driven by a square pulse of duration \SI{8}{\pico\second} with the relevant Rabi frequencies, scaled by $\Omega_{0}$ and $\alpha$. The evolution of all state populations was computed over \SI{1}{\nano\second} using Eq.~\eqref{lind_equation}. 

To compare with experimental transmission traces, modelled populations were converted to the relative transmission change
\[
\frac{\Delta \tau}{\tau} = \frac{\tau(t)-\tau_{\text{ref}}}{\tau_{\text{ref}}}=\frac{\tau(t)}{\tau_{\text{ref}}}-1
\]
where $\tau_{\text{ref}}$ is the transmission of a reference probe with no pump present and the instantaneous transmission follows the Beer-Lambert law:
\[
\tau(t) = \tau_{0} e^{-A(t) }
\]
Here $\tau_{0}$ denotes the baseline transmission away from resonance, accounting only for external factors such as reflection from the sample surfaces. The absorbance is 
\[ A(t)=\sum_{g\in \{1\Gamma_{8}^{+}\}, \,e\in \{1\Gamma_{6}^{-},\,1\Gamma_{7}^{-}\}} \left[ N_{g}(t)-N_{e}(t) \right] \sigma_{eg} d
\]
which is a weighted sum of cross-sections for individual transitions between $g \text{ and } e$, $d$ is the sample thickness and $N_{g,e}$ is the density of atoms in state $g,e$. The individual cross-sections and dipole moments are related by 
\[ 
\sigma_{eg} =\kappa \omega_{eg} |Q_{eg}|^2
\] 
where  $ \omega_{eg} $ is the transition frequency, $Q_{eg}$ is the transition dipole moment from the previous section and $\kappa$ is a proportionality constant that depends only on fundamental constants \cite{Hilborn1982}. 
 Taking $w_{eg}\approx\omega_\text{laser}$, and the dimensionless product $\kappa \omega_\text{laser}\left|D_0\right|^{2} Nd=\lambda $, we have
\[
\tau(t)=\tau_{0}e^{- \tilde{A}(t)\lambda}
\]
where $\tilde{A}$ is a normalised absorbance
\[
\tilde{A}(t)=\sum_{g\in \{1\Gamma_{8}^{+}\}, \, e\in \{1\Gamma_{6}^{-},\,1\Gamma_{7}^{-}\}} \left[ \frac{N_{g}(t)-N_{e}(t)}N \right]\left|\frac{Q_{eg}}{D_0}\right|^2
\]
The equilibrium transmission, where the excited states are empty and the four ground states are equally populated can be used to obtain $\lambda$ through
\[
\tau_\text{ref}=\tau_{0}e^{-\tilde{A}_{\text{ref}}\lambda}
\quad \text{where} \quad
\tilde{A}_\text{ref}=\sum_{g\in \{1\Gamma_{8}^{+}\}, \,e\in \{1\Gamma_{6}^{-},\,1\Gamma_{7}^{-}\}} \frac14\left|\frac{Q_{eg}}{D_0}\right|^2.
\]

The fitting procedure was as follows: the population transients $N_{g,e}(t)$ were calculated using free parameters $\Omega_0$, $\alpha$, $\eta$, and $\xi$. With $\lambda$ and $\tau_0$ fixed, we calculated $\Delta \tau/\tau$ for SCP, OCP and PCP and produced a fit to all the data simultaneously with a single global parameter set. This remarkably simple system resulted in very accurate fits, displayed in the main manuscript. Deviations of the data from the model are likely due to the imperfect circular polarisation, the small difference in the angle of incidence between pump and probe, and crystal orientation uncertainty.

\section*{Supplementary Note 2: Valence Band Effects}

It was important to understand whether the long transient is affected by the valence band, as large pump powers can lead to multi-photon ionisation which is detrimental to spin preparation. Ionised carriers create extra transients in the linear-linear pump-probe \cite{vinh2008, Greenland2010, Vinh2013, Saeedi2018}. 

\begin{figure}[htp]
\centering
\includegraphics[width=0.88\linewidth]{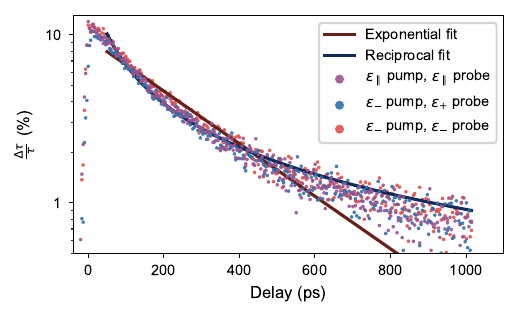}
\caption{
\textbf{Pump-probe directly into the valence band.} Several polarisation combinations were measured at $\sim$ \SI{2.9}{\kelvin} with pulse energy $\sim$ \SI{68}{\nano\joule}, all producing similar transients (photon energy \SI{11.7}{\tera\hertz}). The exponential fit estimated a lifetime of \SI{278}{\pico\second}, much longer than the orbital $T_{1}$ lifetimes of the excited states, but much shorter than the spin $T_{1}$ lifetimes.
}
\label{fig:band_pumping}
\end{figure}

A \SI{11.7}{\tera\hertz} laser exceeding the binding energy excited bound holes directly into the valence band. These unbound holes induced negligible dichroism in the probe (Fig. \ref{fig:band_pumping}). All polarisation combinations yielded the same decay shape: appearing to be mostly reciprocal (as expected for band recombination \cite{vinh2008, Vinh2013}) modulated slightly towards exponential decay, likely due to the competing exponential cascade mechanism decay. 

\begin{figure}[htp]
\centering
\includegraphics[width=1.0\linewidth]{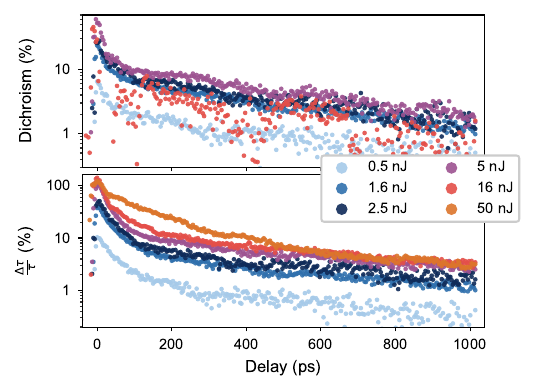}

\caption{
\textbf{Dependence of $1\Gamma_{6}^{-} + 1\Gamma_{7}^{-}$ decay on pump energy.} All measurements were performed at $\sim$ \SI{2.7}{\kelvin}. \textit{Top}: dichroism for several powers up to \SI{16}{\nano\joule}. The long time decays of the dichroism do not change significantly, however at very high powers they begin to decrease in amplitude as the holes become ionised. \textit{Bottom}: the transmission change for SCP ($\epsilon_{-}$ pump $\epsilon_{-}$ probe). The fast and slow lifetimes do not change significantly with power, but at \SI{16}{\nano\joule} a new $\sim$ \SI{200}{\pico\second} component begins to appear corresponding to valence band recombination by photo-ionised holes. At \SI{50}{\nano\joule}, this component is significant.
}

\label{fig:power_dichroism_SCP}
\end{figure}

For all simulations and our estimates of the duration for high-fidelity spin initialisation, we ensured that the pump powers used were in a regime where photo-ionisation was negligible. The threshold for accidental photo-ionisation for pumping into $1\Gamma_{6}^{-} + 1\Gamma_{7}^{-}$ was found to be above \SI{5}{\nano\joule} per $\sim$\SI{9}{\pico\second} pulse (Fig. \ref{fig:power_dichroism_SCP}), thus for modelling we remained in the conservative $<$ \SI{3}{\nano\joule} regime.

\section*{Supplementary Note 3: Proposed Qubit Scheme}

Here we gather insights that could contribute to fully realising the potential of silicon optical pumping. By combining strain-engineering with a magnetic field, it may be possible to reap the benefits of long lifetimes and optical pumping simultaneously. Strain applied to the silicon crystal splits the ground state into two Kramers doublets ($\pm1/2$ and $\pm3/2$). The $\pm1/2$ states can shift downwards and form a lower-lying qubit. It is clear \cite{LUDWIG1962223, Dirksen1989, Kobayashi2021} that by strain-splitting the ground state the $T_{1}$ and $T_{2}$ lifetimes of the $\pm1/2$ qubit can become very long as the single phonon spin-lattice relaxation is suppressed within a Kramers pair. In Ref.~\cite{Kobayashi2021} a magnetic field along $[110]$ and roughly uniform biaxial strain along the $[100]$ and $[010]$ axes lead to \SI{10}{\milli\second} coherence times at \SI{25}{\milli\kelvin} for the newly-formed $\pm1/2$ qubit. Their qubit $T_{1}$ lifetime scales quadratically with strain, and they claim that further strain engineering and a better choice of magnetic field direction would improve $T_{1}$ and coherence time. If both strain and field is applied, many more possibilities for optical pumping pathways appear, including along both $\langle111\rangle$ and $\langle100\rangle$ \cite{Villeret1991}. The new ground state doublet $\pm1/2$ can be coupled to $\pm1/2$ states belonging to $\Gamma_{6}$ and $\Gamma_{7}$, or other higher-lying $\Gamma_{8}$ states that have also strain-split into $\Gamma_{6}$ and $\Gamma_{7}$ Kramers pairs, leaving one ground state dark.

\begin{figure}[htp]
\centering
\includegraphics[width = 1.0\linewidth]{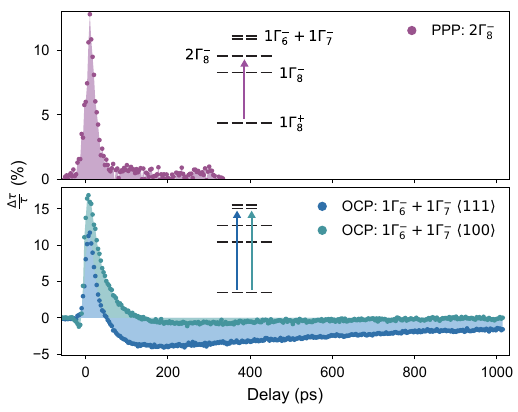}

\caption{
\textbf{Short $2\Gamma_{8}^{-}$ lifetime and crystal orientation dependence.} \textit{Bottom}: the `linear-linear' pump probe (PPP) for the $2\Gamma_{8}^{-}$ state (photon energy \SI{8.3}{\tera\hertz}, sample temperature $\sim$ \SI{2.7}{\kelvin}, pulse energy $\sim$ \SI{4.6}{\nano\joule}) showing a fast, full recovery with orbital $T_{1} = $ \SI{15}{\pico\second}, very similar to results in the literature. \textit{Bottom}: $1\Gamma_{6}^{-} + 1\Gamma_{7}^{-}$ OCP pump-probe for different crystal orientations under similar conditions. The selection rule tables in \cite{Bhattacharjee1972} clearly show a strong dependence on crystal orientation, implying zero dark states when exciting into the $1\Gamma_{6}^{-} + 1\Gamma_{7}^{-}$ manifold along the $\langle 100 \rangle$ axis. This manifested as a large suppression in the initialisation, and a smaller change in absorption when compared to $\langle 111 \rangle$ excitation. For $\langle111\rangle$, the sample temperature was $\sim$ \SI{2.9}{\kelvin} and the pump pulse power was $\sim$ \SI{2.9}{\nano\joule}. For $\langle100\rangle$ (see Methods), the sample temperature was $\sim$ \SI{2.7}{\kelvin} and the pump pulse power was $\sim$ \SI{1.6}{\nano\joule}. 
}

\label{fig:linlin2gamma}
\end{figure}

There are two main limitations to the pulse power and length in the current initialisation scheme: the (\textbf{a}) orbital decay back into the ground states, and (\textbf{b}) photoionisation into the valence band. In particular, for unstrained silicon, optical pumping via $1\Gamma_{6}^{-} + 1\Gamma_{7}^{-}$ to achieve $>99$\% initialisation requires on the order of 1 nanosecond because the decay times of the excited states are tens of picoseconds (effective \SI{36}{\pico\second} decay according to our fits, 56 and \SI{14}{\pico\second} for $1\Gamma_{6}$ and $1\Gamma_{7}$ respectively in older data \cite{Saeedi2018}). Ionisation into the valence band also sets in at pump energies above \SI{5}{\nano\joule} per $\sim$\SI{9}{\pico\second} pulse. Switching to the $2\Gamma_{8}^{-}$ state in boron improves on (\textbf{a}) and (\textbf{b}) simultaneously. We measured it to have an orbital decay lifetime $T_{1} \approx 15.0\pm0.8$\SI{}{\pico\second} (Fig. \ref{fig:linlin2gamma}) almost perfectly matching previous results \cite{Saeedi2018}. It has a very large cross section \cite{Pajot2010}, 
reducing the relative impact of photoionisation. In our simulations outlined below, excitation to the $2\Gamma_{8}^{-}$ sub-levels assumed a \SI{2.9}{\nano\joule} pump pulse; however we performed pump-probe measurements at significantly more power and saw little sign of ionisation (Fig. \ref{fig:linlin2gamma})
, suggesting that we are in a safe, conservative region for modelling.

Simplifying our strained model into the ground state $\pm1/2$ pair and the excited $\pm1/2$ pair originating from the unstrained $2\Gamma_{8}^{-}$ quartet, we simulated the effect that a continuous beam would have on the four populations, assuming that the phonon-mediated decay rates are not significantly affected by the relatively small shifts. Simulations were performed using the strained spin lifetimes taken from various published values: Ref. \cite{Kobayashi2021} ($T_{1} = $ $\sim$ \SI{5}{\ms} at $\sim$ \SI{25}{\milli\kelvin} with \SI{30}{\mega\pascal} stress), Ref. \cite{Dirksen1989} ($T_{1} =$ $\sim$ \SI{4.9}{\ms} at $\sim$ \SI{3}{\kelvin} with \SI{750}{\mega\pascal} stress) and Ref. \cite{LUDWIG1962223} ($T_{1} =$ $\sim$ \SI{3.7}{\ms} at $\sim$ \SI{1.35}{\kelvin} with 490$\pm$\SI{30}{\mega\pascal} stress). The result was that for all lifetimes and temperatures, the target dark state population surpassed 99\% in \SI{245}{\pico\second} (Fig. 5 in the main text). 

Our measurement in Fig. \ref{fig:linlin2gamma} of the orbital $T_{1} \approx 15.0\pm0.8$\SI{}{\pico\second} for excited $2\Gamma_{8}^{-}$ states was likely limited by our pulse duration, which was around 11-\SI{12}{\pico\second}. In \cite{Saeedi2018} they argue that a much shorter pulse duration reveals a more accurate $T_{1}$ of \SI{6.3}{\pico\second} which matches lifetimes expected from linewidths in ultra-pure isotopically purified silicon doped with boron \cite{Steger2009}. With this lifetime in our model, 99\% spin initialisation can be reached in \SI{124}{\pico\second}, still using conservative pump energies. Other acceptors, such as indium or aluminium, have orbital lifetimes on the order of a just a few picoseconds, and would likely lead to initialisation times of less than \SI{100}{\pico\second}. 

\section*{Supplementary Note 4: Temperature Shift}

In Fig. 3b in the main manuscript we plot our THz-measured spin lifetimes. The temperatures were for a thermometer mounted on the same cold finger as the sample, displaying \SI{2.9}{\kelvin} to \SI{11}{\kelvin}. However, unlike the sample, the thermometer was not exposed to the beam or background room-temperature radiation. This causes the reading of the thermometer to diverge from the true sample temperature on cooling, and is always a problem for optical measurements when the sample is exposed to ambient thermal radiation via a cryostat window. In an effort to account for this, we measured a temperature-dependent property of a similar sample using FTIR spectroscopy in a cryostat (similar to that used for the time-resolved experiment) and a dilution fridge with improved shielding from background radiation and a much lower cold finger temperature (\SI{0.8}{\kelvin}). We decided to measure the central frequency shift of the 1s(A$_{1}$) $\rightarrow$ 2p$_{\pm}$ transition as a function of temperature for phosphorus donors in germanium, as they are very sensitive to temperature changes at low temperatures. While not the same semiconductor used in our experiments, germanium and silicon have similar thermal conductivities and the thermal bottleneck is assumed to be the silver-paste mounting, which was done in the same way for both.

We measured the central frequency positions (data unpublished) as a function of temperature for both cryostat and dilution fridge. Assuming the dilution fridge -- able to push far below the base temperature of the $^4$He cryostat used at FELIX -- is the more faithful measurement, we extracted an empirical correction factor $T_\text{shift} = 9.8/T_\text{measured}$ which we then used to shift our measured temperatures for Fig. 3c in the main manuscript. 

\bibliography{bibliography}
